\RequirePackage[l2tabu,orthodox]{nag}
\documentclass[11pt,letterpaper]{article}
\usepackage[T1]{fontenc}
\usepackage[utf8]{inputenc}
\usepackage{crimson}
\usepackage{helvet}
\usepackage[strict,autostyle]{csquotes}
\usepackage[USenglish]{babel}
\usepackage{microtype}
\usepackage{authblk}
\usepackage{booktabs}
\usepackage{caption}
\usepackage{endnotes}
\usepackage{geometry}
\usepackage{graphicx}
\usepackage{hyperref}
\usepackage{natbib}
\usepackage{rotating}
\usepackage{setspace}
\usepackage{titlesec}
\usepackage{url}

\graphicspath{{./figures/}}

\titleformat{\section}{\normalfont\sffamily\large\bfseries}{\thesection.}{0.3em}{}
\titleformat{\subsection}{\normalfont\sffamily\small\bfseries}{\thesubsection.}{0.3em}{}

\makeatletter
\let\@fnsymbol\@arabic
\makeatother

\hypersetup{
	pdfauthor={Geoff Boeing and Dani Arribas-Bel},
	pdftitle={GIS and Computational Notebooks},
	pdfsubject={GIS and Computational Notebooks},
	pdfkeywords={GIS, Computational Notebook, Scientific Computing, Jupyter notebook, Open Science, Open Source},
	pdffitwindow=true,
	breaklinks=true,
	colorlinks=false,
	pdfborder={0 0 0}}

\title{GIS and Computational Notebooks\footnote{\textbf{Citation:} Boeing, G. and D. Arribas-Bel. 2021. \enquote{GIS and Computational Notebooks.} In: \textit{The Geographic Information Science \& Technology Body of Knowledge}, edited by J.~P. Wilson. Ithaca, NY: University Consortium for Geographic Information Science.}}

\author{Geoff Boeing\thanks{Department of Urban Planning and Spatial Analysis, University of Southern California}~ and Dani Arribas-Bel\thanks{Department of Geography and Planning, University of Liverpool}}

\date{}

\begin{document}

\maketitle
\makeatletter
\let\@fnsymbol\@fnsymbol
\makeatother

\begin{abstract}
Researchers and practitioners across many disciplines have recently adopted computational notebooks to develop, document, and share their scientific workflows---and the GIS community is no exception. This chapter introduces computational notebooks in the geographical context. It begins by explaining the computational paradigm and philosophy that underlie notebooks. Next it unpacks their architecture to illustrate a notebook user's typical workflow. Then it discusses the main benefits notebooks offer GIS researchers and practitioners, including better integration with modern software, more natural access to new forms of data, and better alignment with the principles and benefits of open science. In this context, it identifies notebooks as the \enquote{glue} that binds together a broader ecosystem of open source packages and transferable platforms for computational geography. The chapter concludes with a brief illustration of using notebooks for a set of basic GIS operations. Compared to traditional desktop GIS, notebooks can make spatial analysis more nimble, extensible, and reproducible and have thus evolved into an important component of the geospatial science toolkit.
\vspace{1cm}
\end{abstract}

\section{Introduction}

A computational notebook is a computer file that contains code, output, images, and narrative text woven together. Notebooks allow users to consolidate their analytics workflows, blending code, documentation, and results into a single reproducible and distributable file. They also enable interactive computing in the literate programming paradigm. By demonstrating the relevance of these concepts---and this approach---to both research and teaching, this chapter positions computational notebooks as a key emerging tool in the GIS landscape and discusses their advantages for geospatial analysts.

Today many geospatial scholars and practitioners use notebooks to load, clean, filter, analyze, visualize, and model spatial data, as well as to share their workflows and findings with peers and broader audiences. Recent years have witnessed rapid adoption of computational notebooks among data scientists and instructors across disciplines like biology, astronomy, economics, and geography \citep{perkel_why_2018}. These tools have also grown ubiquitous in industry as companies like Google and Airbnb have adopted them for analytics work. To understand this methodological shift, we must consider notebooks' capabilities and the value they create for analysts, researchers, teachers, and students.

\section{How Computational Notebooks Work}

\subsection{The Paradigm}

Scientific researchers and analysts historically used lab notebooks to record their workflow's questions, hypotheses, data, models, results, and all the various analytical decisions made along the way. This was critical for organizing research activities and documenting all the \enquote{whats,} \enquote{whys,} and \enquote{hows} of the scientific process for recollection, replication, and re-analysis. But as graphical user interface (GUI) based analytics software became more common in the late 20\textsuperscript{th} century, many of these details were lost to point-and-click software and the rationales underpinning analytical decisions became obfuscated.

To digitally emulate the benefits of traditional lab notebooks, Mathematica first developed the computational notebook interface in the 1980s as a closed source commercial tool for scientists \citep{somers_scientific_2018}. During the 2010s, open source notebook development, spearheaded by the Jupyter project, expanded throughout research and practice by supporting popular languages in the open source and open science communities such as Python, R, and Julia. Today, computational notebooks mimic traditional lab notebooks digitally and enhance them through two computational paradigms: literate programming and interactive computing.

To explain these two paradigms, we can contrast them with a traditional computer program which consists of lines of code and optional inline comments and is executed linearly from beginning to end. In literate programming, a computer program instead consists of both code and natural language narratives woven together to explain and document the logic of the program \citep{knuth_literate_1992}. In interactive computing, a computer program interacts in real-time with its user and these interactions shape its execution flow \citep{perez_ipython:_2007}. Thus, while a traditional program runs through its code in order, line-by-line, a computational notebook can be executed nonlinearly and can include natural language narratives documenting and explaining each \enquote{chunk} of code alongside its results and output.

Many different kinds of computational notebooks exist today for many different programming languages. It is important to note that there are both general characteristics of notebooks and, separately, individual implementations' own sets of features and elements. Some notebook implementations emphasize code with annotations, as discussed above, while others are more plain-text oriented---but nearly all of these various implementations share a set of common features. As the Jupyter notebook has become the most prominent, we will focus on it as an illustrative example.

\begin{figure*}[tbp]
	\centering
	\includegraphics[width=0.8\textwidth]{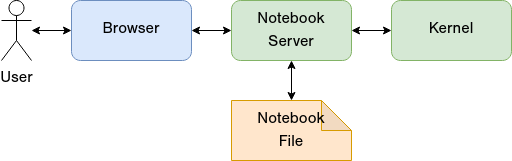}
	\caption{Architecture of the Jupyter notebook computing environment.}
	\label{fig:notebook_architecture}
\end{figure*}

\subsection{Notebook Architecture}

A Jupyter notebook's architecture comprises four components: a web browser, a notebook server, a notebook file, and a kernel \citep{kluyver_jupyter_2016}. As illustrated in Figure \ref{fig:notebook_architecture}, the user relies on a web browser to interact with the notebook server, which reads the notebook file and returns it to the browser to render as a web page. This user interface rendered in the web page includes individual cells---sections of either code or formatted text---in which the user can interactively type. A code cell can be executed by the user: the browser sends the code in a cell to the server, which routes it to the kernel, which runs the code and returns the result to the browser via the notebook server. The browser renders this result inline in the notebook beneath the code cell. Thus, computational output such as individual calculations, tables, or figures appears beside the code that generated it.

Computational notebooks save, store, and present the resulting output of their code within the notebook itself. By integrating these elements, they blend what the code is (code cells) with what the code does (narrative) and with what the code produces (output). The notebook becomes a \enquote{one-stop shop} that allows the user to focus what the code is doing and why and how it is doing it (à la literate programming), as well as to explore and tinker with it while immediately seeing how those changes translate into different results (à la interactive computing).

Another important feature is notebooks' distributed architecture. The notebook server could be running on your own computer, in a server room down the hall, or even on the other side of the world---yet you have easy access to it through any web browser. The server itself is language-agnostic, but notebook kernels are language-specific. Hundreds of Jupyter kernels exist for dozens of different programming languages. The most popular languages for data science and spatial analysis are all supported, including Python, R, and Julia.

\begin{figure*}[tbp]
	\centering
	\includegraphics[width=1\textwidth]{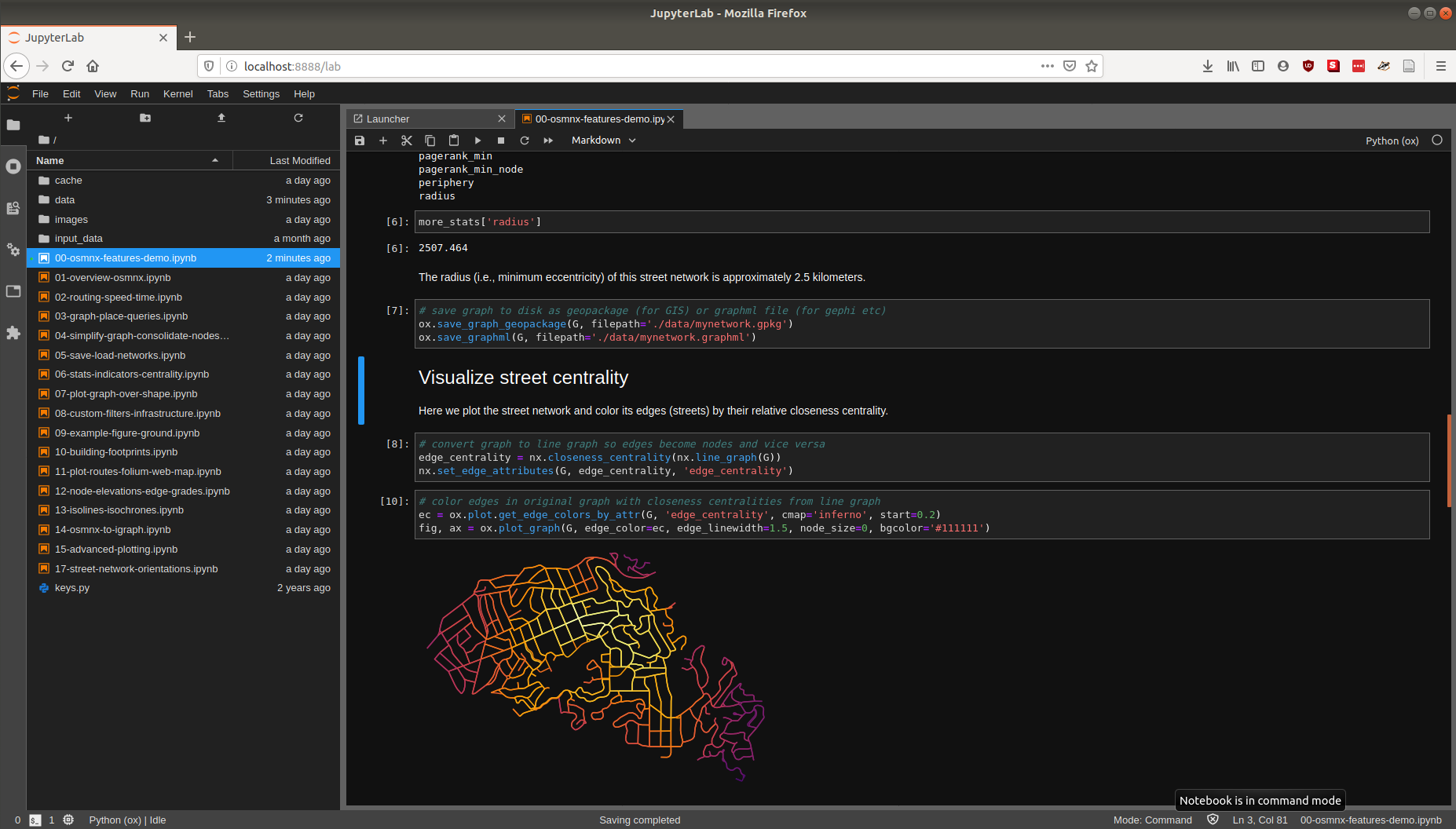}
	\caption{The JupyterLab notebook interface.}
	\label{fig:jupyterlab_interface}
\end{figure*}

\subsection{Notebook Usage}

In the Jupyter ecosystem, the user interface in the web browser is called JupyterLab. JupyterLab shares many common elements with other computational notebooks. It primarily consists of a main work area and a sidebar to browse files and running kernels, as illustrated in Figure \ref{fig:jupyterlab_interface}. The main work area displays the currently open notebooks. Here, notebooks can be created, edited, and executed. Users can add or remove notebook cells, move cells around to reorder their execution, type code or markdown text into cells, or execute one or more cells.

Computational notebooks are increasingly used today in pedagogy, research, and practice. Instructors use them to introduce students to coding and data science because they can show the results of each computation, step-by-step, and explain each new language detail along the way \citep{reades_teaching_2020}. Researchers use them to document, explain, and visualize their questions, hypotheses, data, experiments, and results alongside their analytics code \citep{perkel_why_2018}. Software developers use them to provide visual, narrative usage examples and demonstrations of their software packages for newcomers to learn how they work \citep{boeing_urban_2020}.

All of a computational notebook's code, text, and multimedia content is stored in a single notebook file, so they can easily be shared and distributed, and they work well in collaborative version control systems. Because the specification used to store notebooks is open source and text-based, a wide ecosystem of tools exists to convert between the notebook file format and other formats used in scientific work (e.g., PDF, markdown, LaTeX, Microsoft Word).

\section{Open-Notebook GIS}

Computational notebooks are not specific to GIS. However, the nature of GIS work makes notebooks an essential building block for modern geospatial workflows. Although their domain of use is much wider and spans almost every branch of (interactive) scientific computing, several of their features fit particularly well with existing, well-documented needs in the GIS discipline. If computational notebooks did not exist, the GIS community would have to invent them. Their necessity derives from three core aspects of modern GIS that notebooks address natively: 1) the distributed nature of modern GIS software, 2) the increasingly messy nature of spatial data, and 3) emerging trends in open science.

\subsection{Software Ecosystem}

The GIS software ecosystem has changed dramatically in recent years \citep{arribas-bel_geography_2018}. Until the early 2000s, the desktop had remained the standard platform \citep{gahegan_our_2018}. GIS's desktop-centric paradigm favored all-encompassing software covering as much functionality as possible and exposed it in a simple manner for non-technical audiences. Several GIS industry subdomains still operate in this paradigm today (e.g., local government, military).

However, most of the research community that uses and expands the GIS domain, as well as a nascent industry community around geographic data science, have transitioned to a different model. This new framework employs a distributed and decentralized approach to software production and usage, centered on coding and open source packages. In this modern context, notebooks provide an ideal tool to transparently tie the ecosystem together, because they are built around coding rather than point-and-click desktop GUIs and provide explicit mechanisms to detail the \enquote{whats,} \enquote{whys,} and \enquote{hows} of analytics.

\subsection{Data Ecosystem}

Since the early 2000s, the geospatial data ecosystem has also been radically redefined. Traditional data sources available to geospatial analysts (e.g., decadal censuses, official surveys, limited remote sensing) have been augmented by entirely new kinds of data from sensors such as smartphones, video camera feeds, drones, and nanosatellites. These data are not just \enquote{more} or \enquote{bigger}---they differ fundamentally from traditional data \citep{kitchin_what_2016}. Furthermore, unlike many traditional data sources, they were not explicitly designed for research and rarely come packaged as tidy, organized datasets. Transforming them requires particular effort and attention to render them usable for analytics \citep{harris_more_2017,singleton_geographic_2019}. In this context, notebooks integrate the data science tools required to wrangle and transform these data, empowering analysts to document all the decisions and steps in the process of turning raw inputs into an analysis-ready dataset.

\subsection{GIS and Open Science}

The GIS world has recently engaged in broader discussions concerning open science and reproducible research \citep{brunsdon_quantitative_2016,kedron_reproducibility_2019,kedron_reproducibility_2020}. Recent scandals regarding a lack of transparency and reproducibility have generated attention across the scientific world. In response, reform efforts have coalesced around the notion of open science---which promotes transparent, well-documented, and publicly disseminated research---so third parties can fully understand and replicate the workflows \citep{boeing_right_2020,koster_fueling_2020,wilson_five-star_2020,poorthuis_being_2019,rey_show_2009}.

In GIS, reproducibility should also consider the (too often ignored) underlying technology supporting the research. Although early GIS work was mostly mathematical---and thus easily fully documented in journal articles---the evolution of GUI-based desktop GIS made it difficult to maintain a close correspondence between the numerous operations carried out on the researcher's computer and the final results presented in an article. Computational notebooks present an alternative where transparency and reproducibility are built-in, nudging researchers toward open science.

\enquote{Open-notebook GIS} ties together modern geospatial software and data ecosystems for open science. Notebooks help integrate the modern geospatial software landscape although they are not geospatial software themselves. They help leverage new forms of data whose availability and accessibility relies on modern technologies, such as application programming interfaces (APIs) or distributed databases. In sum, they open up new research pathways while encouraging analytical transparency, documentation, and reproducibility.

\subsection{Notebooks Are Not Enough}

As important as computational notebooks are becoming to modern GIS, they alone are necessary but insufficient. Notebooks can rather be seen as the \enquote{glue} that ties together the various distributed components that make up the modern geospatial scientific stack. This broader landscape of modern GIS---of which computational notebooks are a core component---has two more key building blocks: open source packages and transferable platforms.

First, open source packages are the main vehicle through which scientific software in general (and GIS in particular) is currently distributed. Packages are compilations of code that allow users to reuse and reapply their functionality in a variety of contexts. Open source refers to the license and set of rights that are granted to the user, which typically include examining, modifying, and redistributing the code that makes up the package. Open source packages offer more modular and flexible ways of distributing software than traditional desktop GIS---albeit with an expectation that the user is comfortable writing at least some code. The open source model is more efficient and agile at incorporating new technologies and supporting a wider variety of use cases. In this model, computational notebooks integrate different packages and provide a natural interface that replaces the GUI of traditional desktop GIS.

Second, transferable platforms refer to the broader lower-level set of software and infrastructure required to execute code in a way that is easy to transfer across hardware and users. Modern scientific software stacks are complex and delicate. They rely on many interconnected components that must be installed in a specific way (i.e., version, compiler, etc.) to maintain compatibility with every other piece of the whole. Building and replicating these environments is non-trivial and can hinder reproducibility. The motivation behind transferable platforms is to develop tools and practices that make it easier to distribute these complex stacks in reliable ways. No single standard method for transferable platforms currently exists, but a set of connected components are emerging around package managers like conda and containerization technologies like Docker.

Package managers take care of solving complex package inter-dependencies to ensure the entire stack can function without one package breaking the functionality of another due to version incompatibilities. This situation may seem unlikely or irrelevant, but is in fact a widespread challenge that consumes significant time for data scientists who must ensure packages work well with each other before they can start analyzing data.

Containerization allows users to take a complete software environment that works in one particular context (e.g., a data scientist's latptop), copy it, and then run it elsewhere isolated in a \enquote{container.} Containers ensure all package versions and configurations are the same irrespective of hardware (e.g., laptop, server, or the cloud) and user. In this context, computational notebooks provide the main user interface to a pre-built, fully compartmentalized platform. Containerization makes it easier to distribute transferable platforms in more accessible and user-friendly ways.

\begin{figure*}[!bt]
	\centering
	\includegraphics[width=0.9\textwidth]{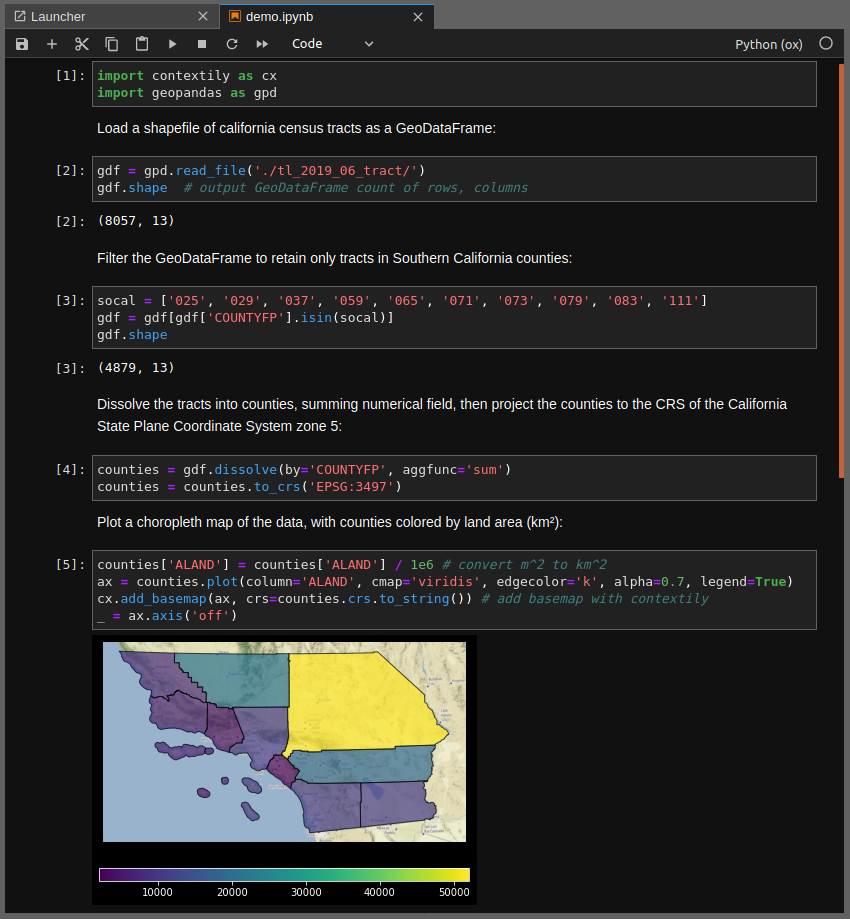}
	\caption{Simple real-world usage example that demonstrates basic GIS in a computational notebook, using JupyterLab and Python.}
	\label{fig:code_demo}
\end{figure*}

\section{Notebooks in Action}

Today, the geographic data science ecosystem is most robust in R (particularly the r-spatial community) and Python, where many packages exist to support spatial analysis and modeling such as \texttt{geopandas} (geospatial data handling), \texttt{PySAL} (advanced spatial analytics and modeling), \texttt{matplotlib} (data visualization), \texttt{OSMnx}  (street network modeling and analytics), \texttt{cenpy} (working with US Census Bureau data), \texttt{contextily} (basemap tiles), \texttt{folium} (web mapping), and many more: see the Additional Resources section for links. These tools allow geospatial data scientists to fully replace traditional desktop GIS software with reproducible, universal analytics workflows in computational notebooks.

Figure \ref{fig:code_demo} presents a real-world usage example using JupyterLab and Python to perform a set of simple GIS operations in a computational notebook. Computational notebooks are not tied to Python---we could perform this example in any programming language that includes such capabilities. Rather, it is the notebook architecture and interactive analytics that matter here.

In cell 1, we import two packages for geospatial analysis and mapping. In cell 2, we read an ESRI shapefile containing all the census tracts in California, turn it into a GeoDataFrame, and output the shape of this resulting GeoDataFrame. We have loaded 8,057 rows (i.e., census tracts) and 13 columns (i.e., variables). Between these two code cells, we see a rendered markdown cell containing a short comment to elucidate the workflow. Markdown provides a simple mechanism for including formatted text such as headings, equations, tables, and hyperlinks. The data in this example are read from a local file, but the code would look similar were we instead to read it directly from a remote location (e.g., a file server such as Amazon Web Services' Landsat 8 satellite imagery catalogue) or to query a web service such as the US Census Bureau API (e.g., with \texttt{cenpy}).

In cell 3, we create a list of all the county IDs in Southern California, then filter the GeoDataFrame to only retain tracts whose county IDs appear in that list. Outputting the shape of the resulting GeoDataFrame reveals that we have retained just 4,879 of the original 8,057 tracts. In cell 4, we dissolve and project the GeoDataFrame. First we aggregate the tracts up to the county level with a standard spatial dissolve operation and sum their numerical attributes to get new county totals. Then we project the GeoDataFrame from its original coordinate reference system (as defined by the shapefile we loaded) to a new one representing the meter-based California State Plane Coordinate System zone 5. Finally, in cell 5, we convert the land area column from square meters to square kilometers, plot a basic choropleth map of the counties colored by land area, and add a basemap.

This notebook can be executed linearly like a script by running it from the top-down, or it can be executed nonlinearly by a user choosing individual cells to run one or multiple times in an arbitrary order. Given the interactive paradigm, a cell or cells can be run once or many times repeatedly, and cells may be run in any order the user desires. Due to this possibly nonlinear flow of execution, it is important to periodically restart the kernel and run all cells to ensure that objects are defined and used in the expected order of operations.

\section{Conclusion}

The open science movement promotes transparency and reproducibility in the workflows underlying data analyses. Traditional desktop GIS's reliance on GUIs makes the software easy to learn and use, but difficult to fully document, understand, and replicate complex spatial analysis projects. Computational notebooks offer an alternative paradigm for scientists, educators, and analysts across disciplines---including quantitative geography. On one hand, notebooks' reliance on code creates a steeper learning curve. On the other hand, they lend themselves much more readily to transparency, reproducibility, documentation, dissemination, and modern data science workflows.

The interested reader is encouraged to explore the Additional Resources section below for \enquote{next steps} getting started with geospatial analytics in computational notebooks.

\setlength{\bibsep}{0.00cm plus 0.05cm}
\bibliographystyle{apalike}
\bibliography{GIS-BoK}

\section*{Definitions}

\begin{itemize}
	\item \textbf{Computational Notebook}: a computer file containing code, output, images, and narrative text woven together.
	\item \textbf{Interactive Scientific Computing}: a method of executing code nonlinearly with interactive user input to develop and run scientific workflows.
	\item \textbf{Jupyter Notebook}: an interactive computing environment comprising a web browser, a notebook server, a notebook file, and a kernel.
	\item \textbf{JupyterLab}: the notebook user interface in the Jupyter ecosystem, rendered by a web browser.
	\item \textbf{Open Science}: a movement to make the workflows and findings of scientific research more accessible, transparent, and reproducible.
	\item \textbf{Open Source Software}: software released under a license that allows users to freely examine, modify, and redistribute its source code.
	\item \textbf{Transferable Platform}: a complete set of underlying software infrastructure to allow code to run consistently and reproducibly across different hardware and users.
\end{itemize}

\section*{Learning Objectives}

\begin{itemize}
    \item Define computational notebooks in general without reference to individual technology platforms.
    \item Explain the difference between the notebook paradigm and traditional desktop GIS.
    \item Explain how to use the JupyterLab user interface.
    \item Describe the utility of computational notebooks in modern GIS analytics.
    \item Express the importance of computational notebooks in open (geospatial) science.
\end{itemize}

\section*{Instructional Assessment Questions}

\begin{itemize}
    \item What is the difference between JupyterLab and a computational notebook?
    \item What are the different components of a Jupyter notebook environment and how do they interrelate?
    \item How do computational notebooks help instructors teach coding?
    \item What are the advantages of using notebooks for geospatial analytics rather than using traditional desktop GIS software?
    \item How do notebooks help researchers make their workflows and findings more transparent and reproducible?
\end{itemize}

\section*{Additional Resources}

\begin{itemize}
    \item \href{https://jupyter.org/}{Project Jupyter}
    \item \href{https://geopandas.org/}{geopandas}
    \item \href{https://pysal.org/}{PySAL}
    \item \href{https://matplotlib.org/}{matplotlib}
    \item \href{https://osmnx.readthedocs.io/}{OSMnx}
    \item \href{https://scitools.org.uk/cartopy/docs/latest/}{cartopy}
    \item \href{https://contextily.readthedocs.io/}{contextily}
    \item \href{https://python-visualization.github.io/folium/}{folium}
    \item \href{https://darribas.org/gds_course}{a course on geographic data science, using notebooks and containerization}
    \item \href{https://github.com/darribas/bok_chapter_notebooks}{a live demonstration of the notebook example in this chapter}
\end{itemize}

\end{document}